\documentclass{elsart}
\usepackage{amssymb}
\usepackage{graphicx}
\usepackage[ulem=normalem]{changes}

\begin{document}

\begin{frontmatter}

\title{Higher-rank discrete symmetries in the IBM.\\
II Octahedral shapes: Dynamical symmetries}

\author[batna]{A.~Bouldjedri},
\author[batna]{S.~Zerguine}
and \author[ganil]{P.~Van~Isacker},

\address[batna]{Department of Physics, PRIMALAB Laboratory, Batna 1 University\\
Route de Biskra, 05000 Batna, Algeria}

\address[ganil]{Grand Acc\'el\'erateur National d'Ions Lourds, CEA/DRF-CNRS/IN2P3\\
Bvd Henri Becquerel, F-14076 Caen, France}

\begin{abstract}
The symmetries of the \mbox{$sdg$-IBM},
the interacting boson model with $s$, $d$ and $g$ bosons,
are studied as regards the occurrence of shapes with octahedral symmetry.
It is shown that no \mbox{$sdg$-IBM} Hamiltonian with a dynamical symmetry
displays in its classical limit an isolated minimum with octahedral shape.
However, a degenerate minimum that includes a shape with octahedral symmetry
can be obtained from a Hamiltonian that is transitional between two limits,
${\rm U}_g(9)\otimes{\rm U}_d(5)$ and ${\rm SO}_{sg}(10)\otimes{\rm U}_d(5)$,
and the conditions for its existence are derived.
An isolated minimum with octahedral shape,
either an octahedron or a cube,
may arise through a modification of two-body interactions between the $g$ bosons.
Comments on the observational consequences of this construction are made.
\end{abstract}

\end{frontmatter}

\section{Introduction}
\label{s_intro}
This paper is the second in the series
initiated with Ref.~\cite{Isacker15}, henceforth referred to as I.
The overall purpose of this series
is the study of nuclear shapes with a higher-rank discrete symmetry,
in particular of the tetrahedral or octahedral type,
in the framework of a variety of interacting boson models.
For the general context of this study we refer the reader to the introduction in I,
of which only the essential points and references are mentioned here.

The paper by Li and Dudek~\cite{Li94}
pointed out the possibility of intrinsic nuclear shapes with a higher-rank discrete symmetry.
Subsequent publications by the Strasbourg group and their collaborators~\cite{Dudek02,Dudek03,Dudek06,Rouvel14}
showed the possible occurrence of tetrahedral, octahedral and icosahedral symmetries
through combinations of deformations of specific multipolarity.
In parallel with these theoretical developments,
a search was initiated for experimental manifestations
of such symmetries in nuclei~\cite{Curien10,Curien11,Jentschel10}.
In addition to these experimental searches cited in I,
a more recent study of this type appeared in 2018~\cite{Dudek18},
formulating criteria for the identification of tetrahedral and/or octahedral symmetries in nuclei.

From the theory side most work related to higher-rank discrete symmetries
has been carried out in the context of mean-field models.
The aim of this series of papers is to address the question
of the possible occurrence of higher-rank discrete symmetries
from a different theoretical perspective.
Our study is carried out in the context of algebraic collective models
inspired by the interacting boson model (IBM)
of Arima and Iachello~\cite{Arima76,Arima78,Arima79},
which proposes a description of quadrupole collective nuclear states
in terms of $s$ and $d$ bosons with angular momentum $\ell=0$ and $\ell=2$.
The application of this idea to collective states of different nature,
notably of octupole and hexadecapole character,
requires the introduction of other bosons~\cite{Iachello87},
in particular $f$ and $g$ bosons with angular momentum $\ell=3$ and $\ell=4$.
Since shapes with tetrahedral symmetry
arise in lowest order through a particular octupole deformation
and those with octahedral symmetry
emanate from a combination of hexadecapole deformations,
the study of such shapes in an algebraic context
requires the introduction of $f$ and $g$ bosons, respectively.

We initiated this program in I with the study of octahedral shapes in the \mbox{$sdg$-IBM}.
We considered the most general rotationally invariant Hamiltonian
with up to two-body interactions between the bosons
and derived the conditions for this Hamiltonian
to have in its classical limit a minimum with an intrinsic shape with octahedral symmetry.
Owing to the general nature of the analysis in I,
only qualitative conclusions could be drawn
with regard to the (non-)existence of such minima in the \mbox{$sdg$-IBM}.
In the present paper we arrive at more concrete conclusions
by considering a subset of all possible Hamiltonians of the \mbox{$sdg$-IBM},
namely those that have a dynamical symmetry and are analytically solvable.
Although none of the symmetry Hamiltonians leads to a stable shape with octahedral symmetry,
a subset can be used to propose a generalization
with additional interactions between the $g$ bosons
that drive the system toward such a shape.

The paper is structured as follows.
To avoid repeatedly referring to equations in I,
we list in Section~\ref{s_sdgibm} formulas from I that are needed in this paper.
In Section~\ref{s_limits} we focus on two dynamical symmetries of the \mbox{$sdg$-IBM}
of particular relevance in our quest for octahedral shapes
and in Section~\ref{s_clas} the classical limit is derived for the Hamiltonian
that is transitional between these two limits.
The central results of this paper are presented in Section~\ref{s_octa},
with a catastrophe analysis of the energy surface
associated with the transitional symmetry Hamiltonian,
and suitable generalizations thereof,
to uncover the existence of minima at stable shapes with octahedral symmetry.
Finally, in Section~\ref{s_conc} the conclusions
of the second paper in this series are summarized.

\section{The \mbox{$sdg$-IBM} and its classical limit}
\label{s_sdgibm}
A boson-number-conserving, rotationally invariant Hamiltonian of the \mbox{$sdg$-IBM}
with up to two-body interactions is of the form
\begin{eqnarray}
\hat H&=&\epsilon_s\hat n_s+
\epsilon_d\hat n_d+
\epsilon_g\hat n_g
\nonumber\\&&
+\sum_{\ell_1\leq\ell_2,\ell'_1\leq\ell'_2,L}
\frac{(-)^Lv^L_{\ell_1\ell_2\ell'_1\ell'_2}}{\sqrt{(1+\delta_{\ell_1\ell_2})(1+\delta_{\ell'_1\ell'_2})}}
[b^\dag_{\ell_1}\times b^\dag_{\ell_2}]^{(L)}\cdot
[\tilde b_{\ell'_2}\times\tilde b_{\ell'_1}]^{(L)},
\label{e_ham}
\end{eqnarray}
where $\hat n_\ell$ is the number operator for the $\ell$ boson,
$\epsilon_\ell$ is the energy of the $\ell$ boson
and $v^L_{\ell_1\ell_2\ell'_1\ell'_2}$ is a boson--boson interaction matrix element.

A geometric understanding of this quantum-mechanical Hamiltonian
is obtained by considering its expectation value in a coherent state.
As discussed in I,
a study of shapes with octahedral symmetry in \mbox{$sdg$-IBM}
requires a coherent state of the form
\begin{equation}
|N;\beta_2,\beta_4,\gamma_2,\gamma_4,\delta_4\rangle=
\sqrt{\frac{1}{N!(1+\beta_2^2+\beta_4^2)^N}}
\Gamma(\beta_2,\beta_4,\gamma_2,\gamma_4,\delta_4)^N|{\rm o}\rangle,
\label{e_coh}
\end{equation}
with
\begin{eqnarray}
\Gamma(\beta_2,\beta_4,\gamma_2,\gamma_4,\delta_4)&=&
s^\dag+
\beta_2\Bigl[\cos\gamma_2d^\dag_0+
\sqrt{\textstyle{\frac 1 2}}\sin\gamma_2(d^\dag_{-2}+d^\dag_{+2})\Bigr]
\label{e_coh1}\\&&+
\beta_4\Bigl[\Bigl(\sqrt{\textstyle{\frac{7}{12}}}\cos\delta_4+
\sqrt{\textstyle{\frac{5}{12}}}\sin\delta_4\cos\gamma_4\Bigr)g^\dag_0
\nonumber\\&&\phantom{+\beta_4\Bigl[}-
\sqrt{\textstyle{\frac 1 2}}\sin\delta_4\sin\gamma_4(g^\dag_{-2}+g^\dag_{+2})
\nonumber\\&&\phantom{+\beta_4\Bigl[}+
\Bigl(\sqrt{\textstyle{\frac{5}{24}}}\cos\delta_4-
\sqrt{\textstyle{\frac{7}{24}}}\sin\delta_4\cos\gamma_4\Bigr)
(g^\dag_{-4}+g^\dag_{+4})\Bigr],
\nonumber
\end{eqnarray}
in terms of the deformation parameters
$\beta_2$, $\beta_4$, $\gamma_2$, $\gamma_4$ and $\delta_4$,
following the convention of Rohozi\'nski and Sobiczewski~\cite{Rohozinski81}.
A shape with octahedral symmetry is obtained for $\beta_2=0$, $\beta_4\neq0$
and $\delta_4=0$ (octahedron) or $\delta_4=\pi$ (cube) with arbitrary $\gamma_4$.
Another solution with octahedral symmetry
exists for $\gamma_4=0$ and $\delta_4=\arccos(1/6)$,
corresponding to a rotated octahedron,
i.e. with the same intrinsic shape as for $\delta_4=0$.
The parameterization introduced in Ref.~\cite{Rohozinski81} therefore
does {\em not} define a unique intrinsic state.

The expectation value of the Hamiltonian~(\ref{e_ham})
in the coherent state~(\ref{e_coh}) (or the classical limit of $\hat H$)
leads to the energy surface
\begin{eqnarray}
\langle\hat H\rangle&\equiv&
E(\beta_2,\beta_4,\gamma_2,\gamma_4,\delta_4)
\nonumber\\&=&
\frac{N(N-1)}{(1+\beta_2^2+\beta_4^2)^2}
\sum_{kl}\beta_2^k\beta_4^l
\left[c'_{kl}+\sum_{ij}c^{ij}_{kl}\cos(i\gamma_2+j\gamma_4)\phi_{kl}^{ij}(\delta_4)\right],
\label{e_climit}
\end{eqnarray}
where $\phi_{kl}^{ij}(\delta_4)$ are trigonometric functions defined in I.
The coefficients $c'_{kl}$ are known
in terms of the scaled single-boson energies $\epsilon'_\ell\equiv\epsilon_\ell/(N-1)$
and the interaction matrix elements $v^L_{\ell_1\ell_2\ell'_1\ell'_2}$,
\begin{eqnarray}&&
\textstyle
c'_{00}={\frac 1 2}v_{ssss}^0+\epsilon'_s,
\nonumber\\&&
\textstyle
c'_{20}=\sqrt{\frac 1 5}v_{ssdd}^0+v_{sdsd}^2+\epsilon'_s+\epsilon'_d,
\nonumber\\&&
\textstyle
c'_{02}={\frac 1 3}v_{ssgg}^0+v_{sgsg}^4+\epsilon'_s+\epsilon'_g,
\nonumber\\&&
\textstyle
c'_{40}=\frac{1}{10}v_{dddd}^0+{\frac 1 7}v_{dddd}^2+\frac{9}{35}v_{dddd}^4+\epsilon'_d,
\nonumber\\&&
\textstyle
c'_{22}=
\frac{1}{\sqrt{45}}v_{ddgg}^0
+\frac{7}{\sqrt{715}}v_{ddgg}^4
+\frac{1}{6}v_{dgdg}^2
+\frac{4}{11}v_{dgdg}^4
+\frac{1}{6}v_{dgdg}^5
+\frac{10}{33}v_{dgdg}^6+\epsilon'_d+\epsilon'_g,
\nonumber\\&&
\textstyle
c'_{04}=
\frac{1}{18}v_{gggg}^0
+\frac{38}{693}v_{gggg}^2
+\frac{89}{1001}v_{gggg}^4
+\frac{62}{495}v_{gggg}^6
+\frac{1129}{6435}v_{gggg}^8+\epsilon'_g.
\label{e_coef1}
\end{eqnarray}
Only a single coefficient $c^{ij}_{kl}$ is needed in the subsequent analysis, {\it viz.}
\begin{equation}
\textstyle
c^{00}_{04}=
-\frac{2}{693}v_{gggg}^2
+\frac{4}{3003}v_{gggg}^4
+\frac{2}{495}v_{gggg}^6
-\frac{16}{6435}v_{gggg}^8,
\label{e_coef2}
\end{equation}
introducing a $\delta_4$ dependence
in the energy surface~(\ref{e_climit}) since
\begin{equation}
\phi^{00}_{04}(\delta_4)=2\cos2\delta_4+17\cos4\delta_4.
\label{e_phi}
\end{equation}
It is assumed in the following that the boson Hamiltonian is Hermitian
and therefore that $v^L_{\ell_1\ell_2\ell'_1\ell'_2}=v^L_{\ell'_1\ell'_2\ell_1\ell_2}$.
With this assumption the expressions for the coefficients $c_{kl}$ and $c^{ij}_{kl}$,
given in Eqs.~(20) and~(21) of I,
are valid with $v^L_{\ell_1\ell_2\cdot\ell'_1\ell'_2}=v^L_{\ell_1\ell_2\ell'_1\ell'_2}$.
Note that for a general Hamiltonian one has
$v^L_{\ell_1\ell_2\cdot\ell'_1\ell'_2}=(v^L_{\ell_1\ell_2\ell'_1\ell'_2}+v^L_{\ell'_1\ell'_2\ell_1\ell_2})/2$,
which corrects by a factor 2 the expression given in I.

\section{The U$_g$(9) and SO$_{sg}$(10) limits}
\label{s_limits}
Since the Hamiltonian of the \mbox{$sdg$-IBM} conserves the total number of bosons,
it can be written in terms of the $(1+5+9)^2=225$ operators $b_{\ell m}^\dag b_{\ell' m'}$.
The 225 operators generate the Lie algebra U(15) with a substructure
that determines the dynamical symmetries of the \mbox{$sdg$-IBM}.

A comprehensive list of the dynamical symmetries of the \mbox{$sdg$-IBM}
is given by De Meyer {\it et al.}~\cite{Meyer86},
and their group-theoretical properties
are extensively discussed by Kota {\it et al.}~\cite{Kota87}.
It is found that the model has seven major dynamical symmetries,
four of strong coupling, SU(3), SU(6), SU(5) and SO(15),
and three of weak coupling,
${\rm U}_s(1)\otimes{\rm U}_{dg}(14)$,
${\rm U}_{sd}(6)\otimes{\rm U}_g(9)$
and ${\rm U}_{sg}(10)\otimes{\rm U}_d(5)$.
The question treated in this paper is
whether any of the dynamical symmetries of the \mbox{$sdg$-IBM}
corresponds to a shape with octahedral symmetry.
In this section a choice of the limits that possibly have such property
is made on the basis of intuitive arguments.
Subsequently, in Sections~\ref{s_clas} and~\ref{s_octa},
the conditions for the existence of a shape with octahedral symmetry
are derived rigorously on the basis of the results obtained in I.

A minimum with octahedral shape
requires mixing of $s$ and $g$ bosons,
so as to induce hexadecapole deformation,
and no or weak mixing of these with the $d$ boson
to ensure zero quadrupole deformation.
These conditions rule out all limits
where $s$, $d$ and $g$ bosons are strongly mixed on an equal footing,
that is, they discard the SU(3), SU(6), SU(5) and SO(15) limits~\cite{Bouldjedri05}.
A strict decoupling of the $s$ and $g$ from the $d$ bosons
is obtained by the reduction
\begin{equation}
\begin{array}{ccccc}
{\rm U}(15)&\supset&{\rm U}_d(5)&\otimes&{\rm U}_{sg}(10)\\
\downarrow&&\downarrow&&\downarrow\\[0mm]
[N]&&n_d&&n_{sg}
\end{array},
\label{e_ds}
\end{equation}
and, furthermore, zero quadrupole deformation
follows from a U(5) classification for the $d$ bosons,
\begin{equation}
\begin{array}{ccccc}
{\rm U}_d(5)&\supset&{\rm SO}_d(5)&\supset&{\rm SO}_d(3)\\
\downarrow&&\downarrow&&\downarrow\\[0mm]
n_d&&\upsilon_d&\nu_d&L_d
\end{array},
\label{e_u5}
\end{equation}
where underneath each algebra the associated quantum number is given.
Specifically, $N$ is the total number of bosons
while $n_\ell$ is the number of $\ell$ bosons
and $n_{\ell\ell'}$ is the number of $\ell$ plus $\ell'$ bosons.
The seniority label associated with an $\ell$ boson is denoted as $\upsilon_\ell$
and corresponds to the number of $\ell$ bosons
not in pairs coupled to angular momentum zero.
Additional (or missing) labels, not associated with any algebra, are indicated with $\nu_\ell$.
Finally, the angular momentum generated by the $\ell$ bosons is denoted as $L_\ell$.

The ${\rm U}_{sg}(10)$ algebra in Eq.~(\ref{e_ds}) allows two classifications of interest.
The first is obtained by eliminating from the generators of ${\rm U}_{sg}(10)$
those that involve the $s$ boson, leading to
\begin{equation}
({\rm I})\qquad
\begin{array}{ccccccc}
{\rm U}_{sg}(10)&\supset&{\rm U}_g(9)&\supset&{\rm SO}_g(9)&\supset&{\rm SO}_g(3)\\
\downarrow&&\downarrow&&\downarrow&&\downarrow\\
n_{sg}&&n_g&&\upsilon_g&\nu_g&L_g
\end{array}.
\label{e_u9}
\end{equation}
In this limit, which for brevity shall be referred to as ${\rm U}_g(9)$ or limit I,
the separate boson numbers $n_s$ and $n_g$ are conserved.
The resulting spectrum is vibrational-like with a spherical ground state
and excited states that correspond to oscillations in the hexadecapole degree of freedom.

The second classification of ${\rm U}_{sg}(10)$ is specified by the following chain of nested algebras:
\begin{equation}
({\rm II})\qquad
\begin{array}{ccccccc}
{\rm U}_{sg}(10)&\supset&{\rm SO}_{sg}(10)&\supset&{\rm SO}_g(9)&\supset&{\rm SO}_g(3)\\
\downarrow&&\downarrow&&\downarrow&&\downarrow\\
n_{sg}&&\upsilon_{sg}&&\upsilon_g&\nu_g&L_g
\end{array},
\label{e_so10}
\end{equation}
which for brevity shall be referred to as ${\rm SO}_{sg}(10)$ or limit II.
The defining feature of the reduction~(\ref{e_so10})
is the appearance of the algebra ${\rm SO}_{sg}(10)$ and its label $\upsilon_{sg}$,
associated with the pairing of $s$ and $g$ bosons.
As is shown in Section~\ref{s_octa},
the ground state in this limit acquires a permanent hexadecapole deformation
and the limit is therefore of interest in our quest for octahedral shapes.
On its own, however, limit II implies degenerate energies of the $s$ and $g$ boson,
and as such it is not realistic.
It is therefore necessary to study a combination of the two limits I and II.

In Sections~\ref{s_clas} and~\ref{s_octa}
we investigate to what extent non-degenerate energies
can be taken for the $s$ and $g$ boson
that still lead to a hexadecapole-deformed minimum
and whether that minimum can have octahedral symmetry.
In the remainder of this section we list some of the properties of limits I and II
that are necessary to carry out this analysis.

The classification of limits I and II can be summarized with the algebraic lattice
\begin{equation}
\begin{array}{ccccc}
{\rm U}(15)&\supset&{\rm U}_d(5)&\otimes&{\rm U}_{sg}(10)\\
&&|&&\swarrow\quad\searrow\\
&&\downarrow&&\quad {\rm U}_g(9)\quad {\rm SO}_{sg}(10)\\
&&{\rm SO}_d(5)&&\searrow\quad\swarrow\\
&&|&&{\rm SO}_g(9)\\
&&\downarrow&&\downarrow\\
&&{\rm SO}_d(3)&&{\rm SO}_g(3)\\
&&\qquad\searrow&&\swarrow\qquad\\
&&&{\rm SO}(3)&
\end{array},
\label{e_lat}
\end{equation}
where SO(3) is associated with the total angular momentum $L$,
which results from the coupling of $L_d$ and $L_g$.
The generators of the different algebras in the lattice~(\ref{e_lat})
are as follows:
\begin{eqnarray}
{\rm U}_d(5)&:\quad&
\{[d^\dag\times\tilde d]^{(\lambda)}_\mu,\lambda=0,\dots,4\},
\nonumber\\
{\rm SO}_d(5)&:\quad&
\{[d^\dag\times\tilde d]^{(\lambda)}_\mu,\lambda=1,3\},
\nonumber\\
{\rm SO}_d(3)&:\quad&
\{\hat L_{d,\mu}\equiv\sqrt{10}[d^\dag\times\tilde d]^{(1)}_\mu\},
\nonumber\\
{\rm U}_{sg}(10)&:\quad&
\{[s^\dag\times\tilde s]^{(0)}_0,[s^\dag\times\tilde g]^{(4)}_\mu,[g^\dag\times\tilde s]^{(4)}_\mu,
[g^\dag\times\tilde g]^{(\lambda)}_\mu,\lambda=0,\dots,8\},
\nonumber\\
{\rm U}_g(9)&:\quad&
\{[g^\dag\times\tilde g]^{(\lambda)}_\mu,\lambda=0,\dots,8\},
\nonumber\\
{\rm SO}_{sg}(10)&:\quad&
\{[s^\dag\times\tilde g+g^\dag\times\tilde s]^{(4)}_\mu,
[g^\dag\times\tilde g]^{(\lambda)}_\mu,\lambda=1,3,5,7\},
\nonumber\\
{\rm SO}_g(9)&:\quad&
\{[g^\dag\times\tilde g]^{(\lambda)}_\mu,\lambda=1,3,5,7\},
\nonumber\\
{\rm SO}_g(3)&:\quad&
\{\hat L_{g,\mu}\equiv\sqrt{60}[g^\dag\times\tilde g]^{(1)}_\mu\},
\nonumber\\
{\rm SO}(3)&:\quad&
\{\hat L_\mu\equiv\hat L_{d,\mu}+\hat L_{g,\mu}\}.
\label{e_gens}
\end{eqnarray}

The linear and quadratic Casimir operators
of the algebras appearing in the lattice~(\ref{e_lat})
can be expressed as follows
in terms of the generators~(\ref{e_gens}): 
\begin{eqnarray}
\hat C_1[{\rm U}(15)]&=&\hat N=\hat n_s+\hat n_d+\hat n_g,
\nonumber\\
\hat C_2[{\rm U}(15)]&=&\hat N(\hat N+14),
\nonumber\\
\hat C_1[{\rm U}_d(5)]&=&\hat n_d,
\nonumber\\
\hat C_2[{\rm U}_d(5)]&=&\hat n_d(\hat n_d+4),
\nonumber\\
\hat C_2[{\rm SO}_d(5)]&=&
2\sum_{\lambda\;{\rm odd}}[d^\dag\times\tilde d]^{(\lambda)}\cdot[d^\dag\times\tilde d]^{(\lambda)},
\nonumber\\
\hat C_2[{\rm SO}_d(3)]&=&\hat L_d\cdot\hat L_d,
\nonumber\\
\hat C_1[{\rm U}_{sg}(10)]&=&\hat n_s+\hat n_g,
\nonumber\\
\hat C_2[{\rm U}_{sg}(10)]&=&(\hat n_s+\hat n_g)(\hat n_s+\hat n_g+9),
\nonumber\\
\hat C_1[{\rm U}_g(9)]&=&\hat n_g,
\nonumber\\
\hat C_2[{\rm U}_g(9)]&=&\hat n_g(\hat n_g+8),
\nonumber\\
\hat C_2[{\rm SO}_{sg}(10)]&=&
[s^\dag\times\tilde g+g^\dag\times\tilde s]^{(4)}\cdot
[s^\dag\times\tilde g+g^\dag\times\tilde s]^{(4)}+
\hat C_2[{\rm SO}_g(9)],
\nonumber\\
\hat C_2[{\rm SO}_g(9)]&=&
2\sum_{\lambda\;{\rm odd}}[g^\dag\times\tilde g]^{(\lambda)}\cdot[g^\dag\times\tilde g]^{(\lambda)},
\nonumber\\
\hat C_2[{\rm SO}_g(3)]&=&\hat L_g\cdot\hat L_g,
\nonumber\\
\hat C_2[{\rm SO}(3)]&=&\hat L\cdot\hat L.
\label{e_cas}
\end{eqnarray}
The expressions for the quadratic Casimir operators of unitary algebras
are not general but are valid in a symmetric irreducible representation.
A rotationally invariant Hamiltonian with up to two-body interactions
can be written in terms of the Casimir operators~(\ref{e_cas}):
\begin{eqnarray}
\hat H_{\rm sym}&=&
\epsilon_d\,\hat n_d+
a_d\,\hat C_2[{\rm U}_d(5)]+
b_d\,\hat C_2[{\rm SO}_d(5)]+
c_d\,\hat C_2[{\rm SO}_d(3)]
\nonumber\\&&+
\epsilon_s\,\hat n_s+\epsilon_g\,\hat n_g+
a_{sg}\,\hat C_2[{\rm U}_{sg}(10)]+
a_g\,\hat C_2[{\rm U}_g(9)]+
b_{sg}\,\hat P_{sg}^\dag\hat P_{sg}
\nonumber\\&&+
b_g\,\hat C_2[{\rm SO}_g(9)]+
c_g\,\hat C_2[{\rm SO}_g(3)]+
c\,\hat C_2[{\rm SO}(3)],
\label{e_hamlat}
\end{eqnarray}
where $\epsilon_\ell$, $a_\ell$, $a_{\ell\ell'}$,
$b_\ell$, $b_{\ell\ell'}$ and $c_\ell$ are parameters.
The quadratic Casimir operator of U(15) is omitted for simplicity
since it gives a constant contribution for a fixed boson number $N=n_s+n_d+n_g$.
Furthermore, it is convenient to define,
instead of the quadratic Casimir operator $\hat C_2[{\rm SO}_{sg}(10)]$,
the combination
\begin{equation}
\hat P^\dag_{sg}\hat P_{sg}=
\hat C_2[{\rm U}_{sg}(10)]-\hat C_1[{\rm U}_{sg}(10)]-\hat C_2[{\rm SO}_{sg}(10)],
\label{e_sfpairing}
\end{equation}
where $\hat P^\dag_{sg}\equiv s^\dag s^\dag-g^\dag\cdot g^\dag$
is the pairing operator for $s$ and $g$ bosons.
The symmetry Hamiltonian~(\ref{e_hamlat}) is less general than Eq.~(\ref{e_ham})
but it is the most general one that can be written
in terms of invariant operators of the lattice~(\ref{e_lat})
and as such it is intermediate between the limits I and II.

The ${\rm U}_g(9)$ limit occurs for $b_{sg}=0$,
leading to the eigenvalues 
\begin{eqnarray}
E_{\rm I}&=&
\epsilon_d\,n_d+
a_d\,n_d(n_d+4)+
b_d\,\upsilon_d(\upsilon_d+3)+
c_d\,L_d(L_d+1)
\nonumber\\&&+
\epsilon_s\,n_s+\epsilon_g\,n_g+
a_{sg}\,n_{sg}(n_{sg}+9)+
a_g\,n_g(n_g+8)+
\nonumber\\&&+
b_g\,\upsilon_g(\upsilon_g+7)+
c_g\,L_g(L_g+1)+
c\,L(L+1).
\label{e_eigu9}
\end{eqnarray}
The ${\rm SO}_{sg}(10)$ limit is attained
for $\epsilon_s=\epsilon_g\equiv\epsilon_{sg}$ and $a_g=0$,
in which case the Hamiltonian's eigenstates have the eigenvalues
\begin{eqnarray}
E_{\rm II}&=&
\epsilon_d\,n_d+
a_d\,n_d(n_d+4)+
b_d\,\upsilon_d(\upsilon_d+3)+
c_d\,L_d(L_d+1)
\nonumber\\&&+
\epsilon_{sg}\,n_{sg}+
a_{sg}\,n_{sg}(n_{sg}+9)+
b_{sg}[n_{sg}(n_{sg}+8)-\upsilon_{sg}(\upsilon_{sg}+8)]
\nonumber\\&&+
b_g\,\upsilon_g(\upsilon_g+7)+
c_g\,L_g(L_g+1)+
c\,L(L+1).
\label{e_eigso10}
\end{eqnarray}
The eigenspectra are then determined
with the help of the necessary branching rules.
The reduction ${\rm U}(15)\supset{\rm U}_d(5)\otimes{\rm U}_{sg}(10)$
implies the relation $N=n_d+n_{sg}$ or the branching rule
\begin{equation}
[N]\mapsto(n_d,n_{sg})=(0,N),(1,N-1),\dots,(N,0).
\label{e_branch}
\end{equation}
The branching rules for the classification~(\ref{e_u5})
are known from the U(5) limit of the \mbox{$sd$-IBM}~\cite{Arima76}
and those for the classifications~(\ref{e_u9}) and~(\ref{e_so10})
can be found in Ref.~\cite{Kota87}.

\begin{figure}
\centering
\includegraphics[width=6.5cm]{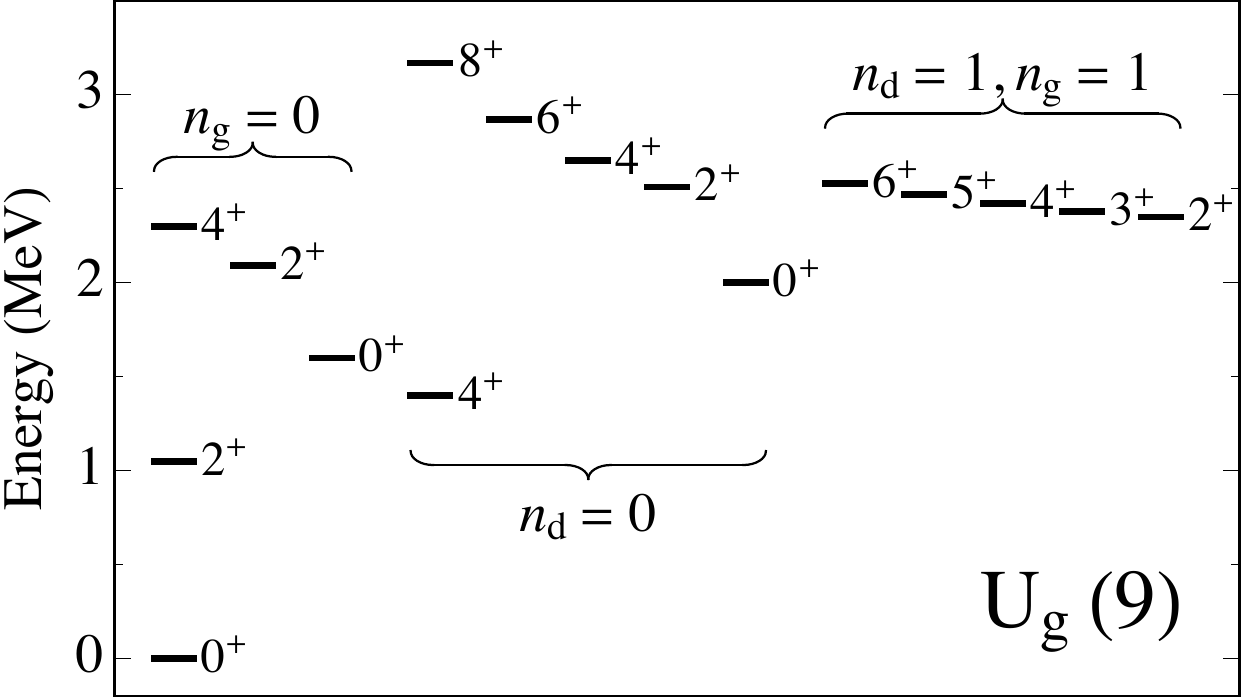}
\includegraphics[width=6.5cm]{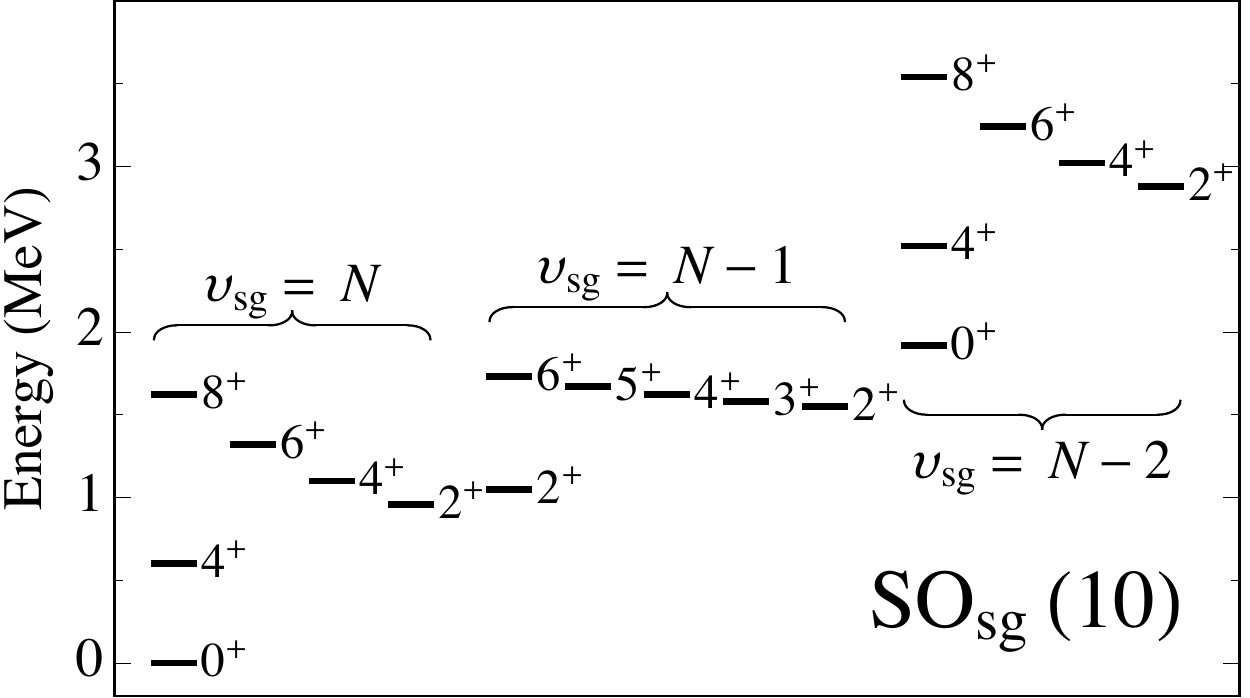}
\caption{Energy spectra in the ${\rm U}_g(9)$ and ${\rm SO}_{sg}(10)$ limits
of the \mbox{$sdg$-IBM} for $N=5$ bosons.
For the ${\rm U}_g(9)$ spectrum the non-zero parameters in the Hamiltonian~(\ref{e_hamlat})
are $\epsilon_d-\epsilon_s=800$, $\epsilon_g-\epsilon_s=1000$,
$b_d=40$, $c_d=10$, $b_g=25$ and $c_g=c=5$~keV.
For the ${\rm SO}_{sg}(10)$ spectrum the non-zero parameters
are $\epsilon_d-\epsilon_s=800$, $\epsilon_g-\epsilon_s=0$,
$b_d=40$, $c_d=10$, $b_{sg}=60$, $b_g=50$ and $c_g=c=5$~keV.}
\label{f_limits}
\end{figure}
Typical energy spectra in the ${\rm U}_g(9)$ and ${\rm SO}_{sg}(10)$ limits
are shown in Fig.~\ref{f_limits}.
The ${\rm U}_g(9)$ spectrum displays quadrupole- and hexadecapole-phonon multiplets
characterized by a fixed number of $d$  and $g$ bosons.
The multiplets are further structured by a seniority quantum number:
for example, the $n_d=2$ multiplet has $\upsilon_d=2$ except for the $0^+$ level,
which has $\upsilon_d=0$,
and similarly for the $n_g=2$ multiplet and the $\upsilon_g$ seniority.
Also combined quadrupole--hexadecapole multiplets occur in the spectrum.
The ${\rm SO}_{sg}(10)$ spectrum
contains sets of levels with $\upsilon_{sg}=N,N-2,\dots$ (for $n_d=0$)
with $\upsilon_{sg}=N-1,N-3,\dots$ (for $n_d=1$) etc.,
and $\upsilon_{sg}=N$ levels are lowest in energy due the repulsive $sg$-pairing.
Multiplets characterized by the seniority quantum number $\upsilon_g=0,1,\dots$
occur within each ${\rm SO}_{sg}(10)$ multiplet.
Note that in this limit, unless $b_{sg}$ is small, the first-excited state has $J^\pi=4^+$.
This is a consequence of the unrealistic condition
that the energies of the $s$ and $g$ bosons are degenerate, $\epsilon_s=\epsilon_g$.

\section{Classical limit of the symmetry Hamiltonian}
\label{s_clas}
The classical limit of the symmetry Hamiltonian~(\ref{e_hamlat})
can be obtained with the general procedure outlined in I.
First, a conversion to the standard representation~(\ref{e_ham}) is carried out.
A quadratic Casimir operator $\hat C_2(G)$
is given generically as an expansion over the generators, 
\begin{equation}
\hat C_2(G)=
\sum_{\lambda r}a_{\lambda r}
\left(\sum_{\ell_1\ell_2}\alpha^{\lambda r}_{\ell_1\ell_2}
\left(b_{\ell_1}^\dag\times\tilde b_{\ell_2}\right)^{(\lambda)}\right)
\cdot
\left(\sum_{\ell'_1\ell'_2}\alpha^{\lambda r}_{\ell'_1\ell'_2}
\left(b_{\ell'_1}^\dag\times\tilde b_{\ell'_2}\right)^{(\lambda)}\right),
\label{e_casexp}
\end{equation}
with coefficients $\alpha^{\lambda r}_{\ell_1\ell_2}$ and $a_{\lambda r}$
that are specific to each algebra $G$, as given in Eq.~(\ref{e_cas}).
The overall sum in Eq.~(\ref{e_casexp})
is over the multipolarity $\lambda$ of the generators
and over an additional index $r$
to distinguish different generators with the same $\lambda$.
With use of the expansion~(\ref{e_casexp})
one finds the following expression
for the matrix element of $\hat C_2(G)$ between two-boson states~\cite{note1}:
\begin{eqnarray}
\lefteqn{\langle \ell_1\ell_2;L|\hat C_2(G)|\ell'_1\ell'_2;L\rangle}
\nonumber\\&=&
\left[f_{\ell_1}(G)+f_{\ell_2}(G)\right]
\delta_{\ell_1\ell_3}\delta_{\ell_2\ell_4}+
\frac{2(-)^{\ell_2+\ell_3}}{\sqrt{(1+\delta_{\ell_1\ell_2})(1+\delta_{\ell_3\ell_4})}}
\sum_{\lambda r}a_{\lambda r}(2\lambda+1)
\nonumber\\&&\times
\Biggl[\alpha^{\lambda r}_{\ell_1\ell_4}\alpha^{\lambda r}_{\ell_2\ell_3}
\Biggl\{\begin{array}{ccc}
\ell_1&\ell_4&\lambda\\ 
\ell_3&\ell_2&L
\end{array}\Biggr\}+
(-)^L\alpha^{\lambda r}_{\ell_1\ell_3}\alpha^{\lambda r}_{\ell_2\ell_4}
\Biggl\{\begin{array}{ccc}
\ell_1&\ell_3&\lambda\\ 
\ell_4&\ell_2&L
\end{array}
\Biggr\}\Biggr],
\label{e_convert}
\end{eqnarray}
where the symbol between curly brackets is a Racah coefficient~\cite{Talmi93}.
The quantity $f_\ell(G)$ is the expectation value
of the operator $\hat C_2(G)$ between single-boson states,
\begin{equation}
f_\ell(G)\equiv\langle \ell|\hat C_2(G)|\ell\rangle=
\sum_{\lambda r}\frac{2\lambda+1}{2\ell+1}a_{\lambda r}
\sum_{\ell'}(-)^{\ell+\ell'}
\alpha^{\lambda r}_{\ell\ell'}
\alpha^{\lambda r}_{\ell'\ell}.
\label{e_conone}
\end{equation}

The Hamiltonian~(\ref{e_hamlat}) can,
with use of Eqs.~(\ref{e_convert}) and~(\ref{e_conone}),
be converted into its standard representation
on which the classical-limit expression~(\ref{e_climit}) can be applied.
The procedure results in the energy surface
\begin{equation}
E(\beta_2,\beta_4)=
\frac{N(N-1)}{(1+\beta_2^2+\beta_4^2)^2}
\sum_{kl}c'_{kl}\beta_2^k\beta_4^l,
\label{e_climit1}
\end{equation}
where the non-zero coefficients $c'_{kl}$ are
\begin{eqnarray}
c'_{00}&=&
a_{sg}+b_{sg}+\Gamma_s,
\quad
c'_{20}=
\Gamma_s+\Gamma_d,
\quad
c'_{02}=
2a_{sg}-2b_{sg}+\Gamma_s+\Gamma_g,
\nonumber\\
c'_{40}&=&
a_d+\Gamma_d,
\quad
c'_{22}=
\Gamma_d+\Gamma_g,
\quad
c'_{04}=
a_{sg}+b_{sg}+a_g+\Gamma_g,
\label{e_coeflat}
\end{eqnarray}
in terms of the combinations
\begin{eqnarray}
\Gamma_s&\equiv&\frac{1}{N-1}(\epsilon_s+10a_{sg}),
\nonumber\\
\Gamma_d&\equiv&\frac{1}{N-1}(\epsilon_d+5a_d+4b_d+6c_d+6c),
\nonumber\\
\Gamma_g&\equiv&\frac{1}{N-1}(\epsilon_g+10a_{sg}+9a_g+8b_g+20c_g+20c).
\label{e_comb}
\end{eqnarray}
All coefficients $c^{ij}_{kl}$ of Eq.~(\ref{e_climit}) vanish identically
in the classical limit of the symmetry Hamiltonian~(\ref{e_hamlat}).

\section{Octahedral shapes}
\label{s_octa}
From the outset it should be clear that energy surface~(\ref{e_climit1})
cannot have an isolated minimum with octahedral shape
since it is independent of $\gamma_2$, $\gamma_4$ and $\delta_4$.
What can still happen, however, is the occurrence of a minimum
with zero quadrupole and non-zero hexadecapole deformation
($\beta^*_2=0$ and $\beta^*_4\neq0$),
which, given the instability in $\gamma_4$ and $\delta_4$,
{\em includes} a shape with octahedral symmetry.
Therefore, the goal of this section is
to establish the conditions on the parameters in the symmetry Hamiltonian~(\ref{e_hamlat})
such that its classical limit displays a minimum with $\beta^*_2=0$ and $\beta^*_4\neq0$,
and, subsequently, to identify the interactions in the general Hamiltonian~(\ref{e_ham})
that generate a dependence on $\delta_4$,
enabling the formation of an isolated minimum with octahedral shape.
This problem can be investigated with the procedure outlined in I.

According to the analysis of the previous section,
the classical energy~(\ref{e_climit1}) is a two-variable function
$E(\beta_2,\beta_4)$.
The conditions for this energy surface
to have an extremum are
\begin{equation}
\left.\frac{\partial E}{\partial\beta_2}\right|_{p^*}=
\left.\frac{\partial E}{\partial\beta_4}\right|_{p^*}=0,
\label{e_extr}
\end{equation}
where $p^*\equiv(\beta^*_2,\beta^*_4)$
is a short-hand notation for an arbitrary critical point.
Furthermore, a critical point at an extremum
with $\beta^*_2=0$ and $\beta^*_4\neq0$ shall be denoted as $h^*$.
The condition~(\ref{e_extr}) in $\beta_2$ is identically satisfied for $p^*=h^*$
and does not lead to any constraints on the coefficients $c'_{kl}$.
The condition in $\beta_4$ leads to a cubic equation with the solutions
\begin{equation}
\beta_4^*=0,
\qquad
\beta_4^*=\pm\sqrt{\frac{2c'_{00}-c'_{02}}{2c'_{04}-c'_{02}}}.
\label{e_extrb4}
\end{equation}
Only the last solution corresponds to an extremum $h^*$
and implies the following condition on the ratio of coefficients:
\begin{equation}
\frac{2c'_{00}-c'_{02}}{2c'_{04}-c'_{02}}>0.
\label{e_cond1}
\end{equation}

While the condition~(\ref{e_cond1}) is necessary and sufficient
to have an {\em extremum} at $\beta_2^*=0$ and $\beta_4^*\neq0$,
a {\it minimum} at these values implies further constraints.
They are obtained by requiring that the eigenvalues of the stability matrix
[{\it i.e.}, the partial derivatives of $E(\beta_2,\beta_4)$
of second order] are all positive.
Since the off-diagonal element of the stability matrix vanishes for the energy surface~(\ref{e_climit1}),
the existence of a minimum follows from the uncoupled conditions
\begin{equation}
\left.\frac{\partial^2 E}{\partial\beta_2^2}\right|_{h^*}>0,
\qquad
\left.\frac{\partial^2 E}{\partial\beta_4^2}\right|_{h^*}>0,
\label{e_stab1}
\end{equation}
or, in terms of the coefficients $c'_{kl}$ in Eq.~(\ref{e_climit1}),
\begin{eqnarray}
\frac{(2c'_{04}-c'_{02})[2c'_{00}(c'_{22}-2c'_{04})+c'_{02}(c'_{02}-c'_{20}-c'_{22})+2c'_{04}c'_{20}]}
{(c'_{00}-c'_{02}+c'_{04})^2}&>&0,
\nonumber\\
\frac{(2c'_{00}-c'_{02})(2c'_{04}-c'_{02})^3}{(c'_{00}-c'_{02}+c'_{04})^3}&>&0.
\label{e_stab2}
\end{eqnarray}
If we write Eq.~(\ref{e_cond1}) as $A/B>0$,
the second inequality in Eq.~(\ref{e_stab2}) becomes $AB^3/(A+B)^3>0$
and therefore both $A$ and $B$ should be positive,
$A\equiv2c'_{00}-c'_{02}>0$ and $B\equiv2c'_{04}-c'_{02}>0$,
leading to the constraint
\begin{equation}
-4b_{sg}-2a_g<\Gamma_g-\Gamma_s<4b_{sg}.
\label{e_cond2}
\end{equation}
The first inequality in Eq.~(\ref{e_stab2}) can be reduced to
\begin{equation}
2b_{sg}(2\Gamma_d-\Gamma_s-\Gamma_g-4a_{sg}-a_g)
+a_g(\Gamma_d-\Gamma_s-2a_{sg})>0.
\label{e_cond3}
\end{equation}

The conditions~(\ref{e_cond2}) and~(\ref{e_cond3}) are necessary and sufficient
for the energy surface $E(\beta_2,\beta_4)$
to have a minimum at zero quadrupole and non-zero hexadecapole deformation.
To obtain an intuitive understanding of them,
we note that $\Gamma_g-\Gamma_s$, for a reasonable choice of parameters, is positive.
The upper part of the inequality~(\ref{e_cond2}) therefore expresses
the need for $b_{sg}$ to be positive and sufficiently large,
corresponding to a repulsive $sg$-pairing interaction
that puts the configuration with {\em maximal} $sg$ seniority $\upsilon_{sg}=n_{sg}$ at lowest energy.
For $b_{sg}>0$ and $a_g>0$,
the lower part of the inequality~(\ref{e_cond2}) is automatically satisfied.
The condition~(\ref{e_cond3}) is easier to appreciate
if it is assumed that the coefficients in front of the quadratic Casimir operators
of the unitary algebras ${\rm U}_d(5)$, ${\rm U}_g(9)$ and ${\rm U}_{sg}(10)$ vanish,
$a_d=a_{sg}=a_g=0$.
This assumption is justified if anharmonicities are neglected in the various limits.
Given that $b_{sg}>0$, it then follows that
\begin{equation}
2\Gamma_d-\Gamma_s-\Gamma_g>0.
\label{e_cond4}
\end{equation}
In terms of the original parameters in the symmetry Hamiltonian~(\ref{e_hamlat})
(assuming $a_d=a_{sg}=a_g=0$)
the conditions to have a minimum at $\beta^*_2=0$ and $\beta^*_4\neq0$
can be summarized as
\begin{eqnarray}
-4(N-1)b_{sg}<\epsilon_g-\epsilon_s+8b_g+20(c_g+c)&<&4(N-1)b_{sg},
\nonumber\\
2\epsilon_d-\epsilon_s-\epsilon_g+8(b_d-b_g)+4(3c_d-5c_g-2c)&>&0.
\label{e_cond}
\end{eqnarray}

We now ask the question whether two-body interactions
can be added to the symmetry Hamiltonian~(\ref{e_hamlat}),
which lift the $(\gamma_4,\delta_4)$ instability
and create a minimum at $\delta_4=0$, $\delta_4=\arccos(1/6)$ or $\delta_4=\pi$.
To achieve this goal, we recall the result from I
that for a general Hamiltonian of the \mbox{$sdg$-IBM}
an extremum with $\beta_2^*=0$ and $\beta_4^*\neq0$ occurs for
\begin{equation}
\beta_4^*=\pm\sqrt{\frac{2c'_{00}-c'_{02}}{2c'_{04}-c'_{02}+38c^{00}_{04}}}.
\label{e_extrb4g}
\end{equation}
The term in $c^{00}_{04}$ introduces a dependence in $\delta_4$
that may lead to an isolated minimum with octahedral symmetry.
Given the expression~(\ref{e_coef2}) for $c^{00}_{04}$,
this argument suggests adding $g$-boson interactions $v_{gggg}^L$ to the Hamiltonian~(\ref{e_hamlat}).
The classical energy~(\ref{e_climit1}) then becomes a three-variable function
$E(\beta_2,\beta_4,\delta_4)$,
for which the above catastrophe analysis can be repeated.
With these additional two-body interactions
the extremum and stability conditions~(\ref{e_extr}) and~(\ref{e_stab1}) become
\begin{eqnarray}
&&2c'_{00}-c'_{02}>0,
\qquad
2c'_{04}-c'_{02}+38c^{00}_{04}>0,
\qquad
c^{00}_{04}<0,
\label{e_cong1}\\
&&2c'_{00}(c'_{22}-2c'_{04})+c'_{02}(c'_{02}-c'_{20}-c'_{22})+2c'_{04}c'_{20}-38c^{00}_{04}(2c'_{00}-c'_{20})>0.
\nonumber
\end{eqnarray}
For $a_d=a_{sg}=a_g=0$,
these conditions imply the inequalities
\begin{eqnarray}
&&-2b_{sg}-\tilde v<\Gamma_g-\Gamma_s<4b_{sg},
\nonumber\\
&&\bar v\equiv 65v_{gggg}^2-30v_{gggg}^4-91v_{gggg}^6+56v_{gggg}^8>0,
\nonumber\\
&&4b_{sg}(2\Gamma_d-\Gamma_s-\Gamma_g)
+(\Gamma_d-\Gamma_s-2b_{sg})(\tilde v-2b_{sg})>0,
\label{e_cong2}
\end{eqnarray}
where $\tilde v$ is the following linear combination of $g$-boson interaction matrix elements:
\begin{equation}
\textstyle
\tilde v=
\frac{1}{9}v_{gggg}^0
+\frac{98}{429}v_{gggg}^4
+\frac{40}{99}v_{gggg}^6
+\frac{10}{39}v_{gggg}^8.
\label{e_vtilde}
\end{equation}
For the Hamiltonian~(\ref{e_hamlat}) the linear combination $\bar v$ vanishes identically
and the second inequality in Eq.~(\ref{e_cong2}) is not fulfilled.
This expresses the $\delta_4$ independence of the symmetry Hamiltonian
and the fact that its classical limit does not acquire an isolated minimum with octahedral shape.
Furthermore, for the symmetry Hamiltonian one has $\tilde v=2b_{sg}$
and the conditions~(\ref{e_cong2}) reduce to Eq.~(\ref{e_cond}).

There are clearly many ways to find matrix elements $v_{gggg}^L$
that satisfy all conditions~(\ref{e_cong2}) but one way is particularly simple.
Note that the quadrupole matrix element $v_{gggg}^2$
does not appear in the combination $\tilde v$.
By making this matrix element more repulsive,
the second inequality in Eq.~(\ref{e_cong2}) is satisfied
while the other two conditions are not modified
with respect those in Eq.~(\ref{e_cond}) valid for the symmetry Hamiltonian~(\ref{e_hamlat}).
A possible procedure to construct an \mbox{$sdg$-IBM} Hamiltonian
whose classical energy displays a minimum with octahedral shape
is therefore to add to a hexadecapole-deformed symmetry Hamiltonian~(\ref{e_hamlat})
a repulsive $v_{gggg}^2$ interaction.

\begin{figure}
\centering
\includegraphics[width=7cm]{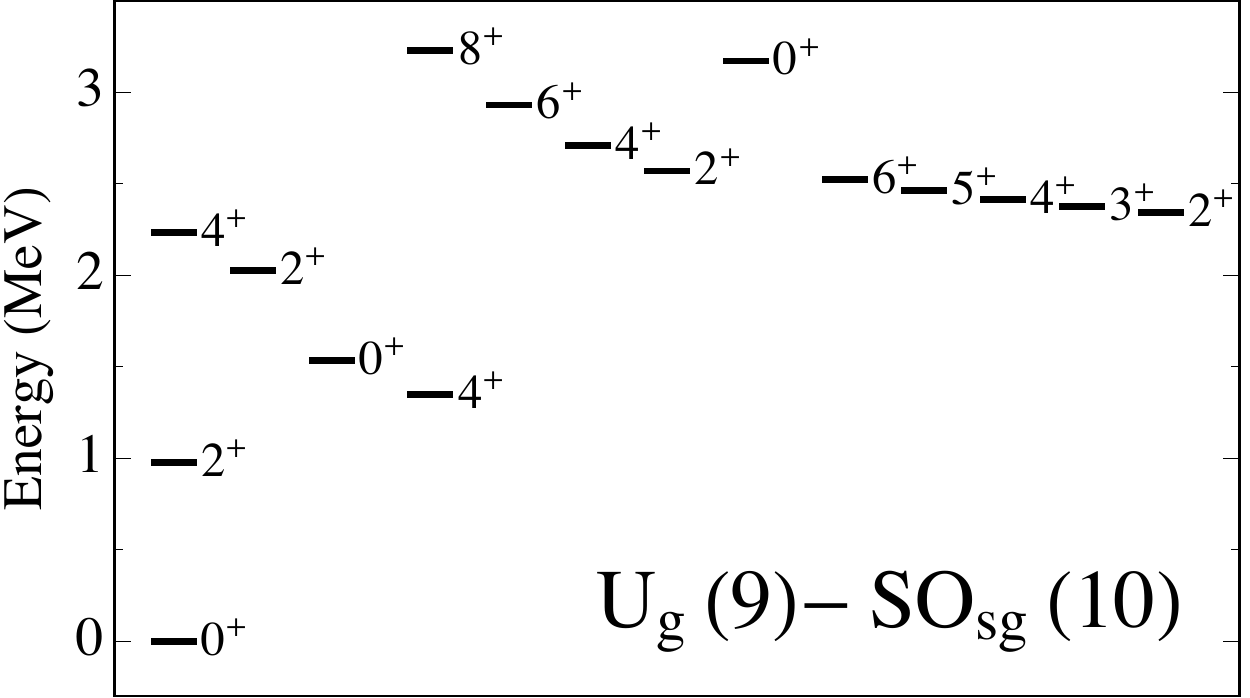}
\caption{Energy spectrum of a ${\rm U}_g(9)$--${\rm SO}_{sg}(10)$ transitional Hamiltonian
of the \mbox{$sdg$-IBM} for $N=5$ bosons.
The non-zero parameters of the Hamiltonian~(\ref{e_hamlat})
are $\epsilon_d-\epsilon_s=1200$, $\epsilon_g-\epsilon_s=1500$,
$b_d=40$, $c_d=10$, $b_{sg}=150$, $b_g=25$ and $c_g=c=5$~keV.}
\label{f_u9so10}
\end{figure}
Let us illustrate this procedure with an example.
The starting point is a ${\rm U}_g(9)$--${\rm SO}_{sg}(10)$ transitional Hamiltonian
associated with the lattice~(\ref{e_lat}),
giving rise to the spectrum shown in Fig.~\ref{f_u9so10}.
Note that the choice of the single-boson energies $\epsilon_\ell$ for this figure
is realistic in the sense that the $g$-boson energy
is higher than that of the $d$ boson.
The $sg$-pairing strength $b_{sg}$,
of which little is known either empirically or microscopically,
is chosen such that a hexadecapole-deformed minimum occurs in the classical limit.
Other parameters in the Hamiltonian~(\ref{e_hamlat}) are of lesser importance
and are chosen as to lift degeneracies in the spectrum.
Note that with this choice of parameters
the resulting spectrum, as shown in Fig.~\ref{f_u9so10},
is rather closer to the ${\rm U}_g(9)$ and than to the ${\rm SO}_{sg}(10)$ limit.

\begin{figure}
\centering
\includegraphics[width=6cm]{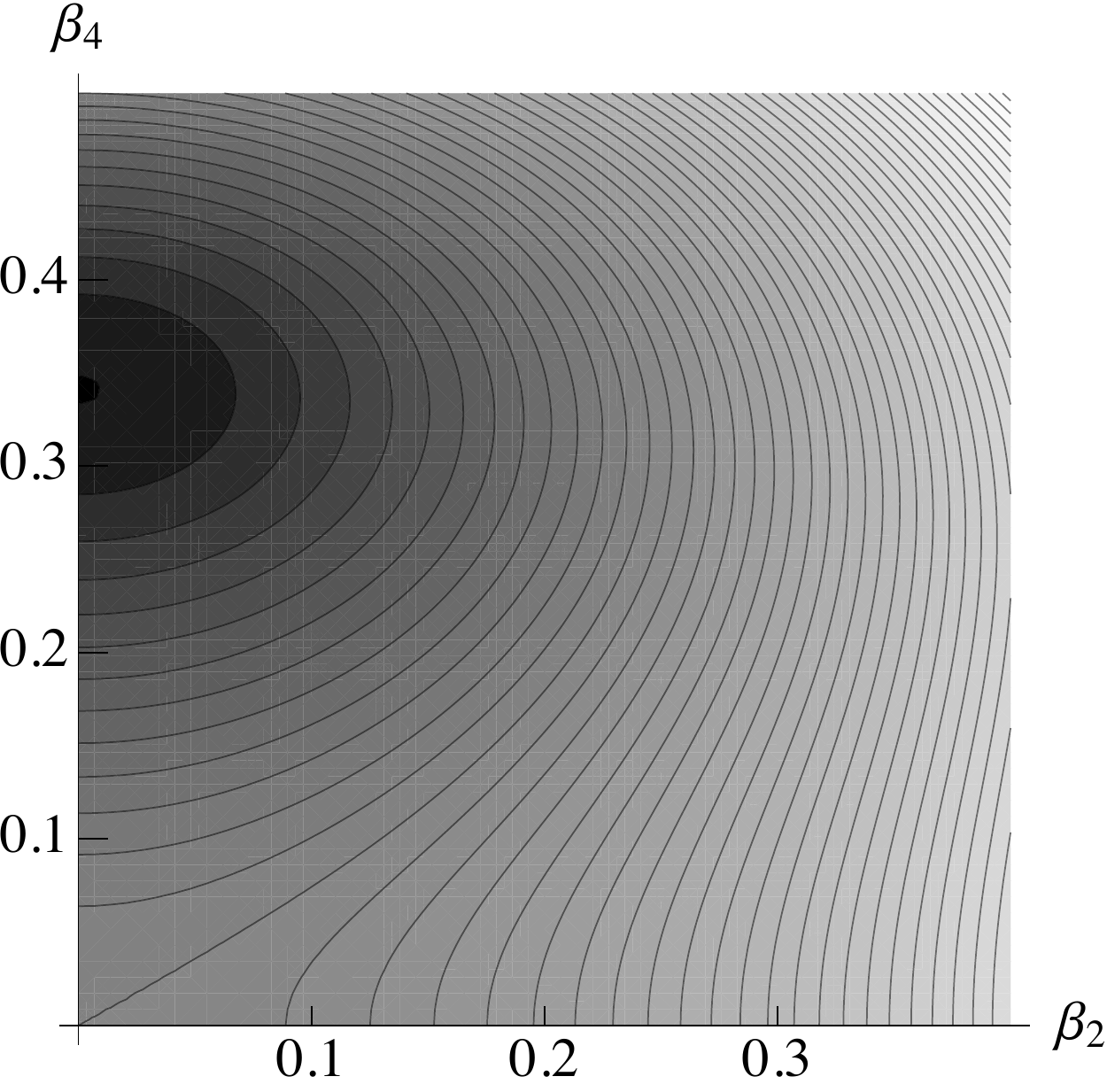}
\caption{The energy surface $E(\beta_2,\beta_4)$
obtained in the classical limit 
of a ${\rm U}_g(9)$--${\rm SO}_{sg}(10)$ transitional Hamiltonian of the \mbox{$sdg$-IBM}.
Parameters of the Hamiltonian~(\ref{e_hamlat})
are given in the caption of Fig.~\ref{f_u9so10}.
Black corresponds to low energies
and the lines indicate changes by 10~keV.}
\label{f_b2b4}
\end{figure}
\begin{figure}
\centering
\includegraphics[width=6cm]{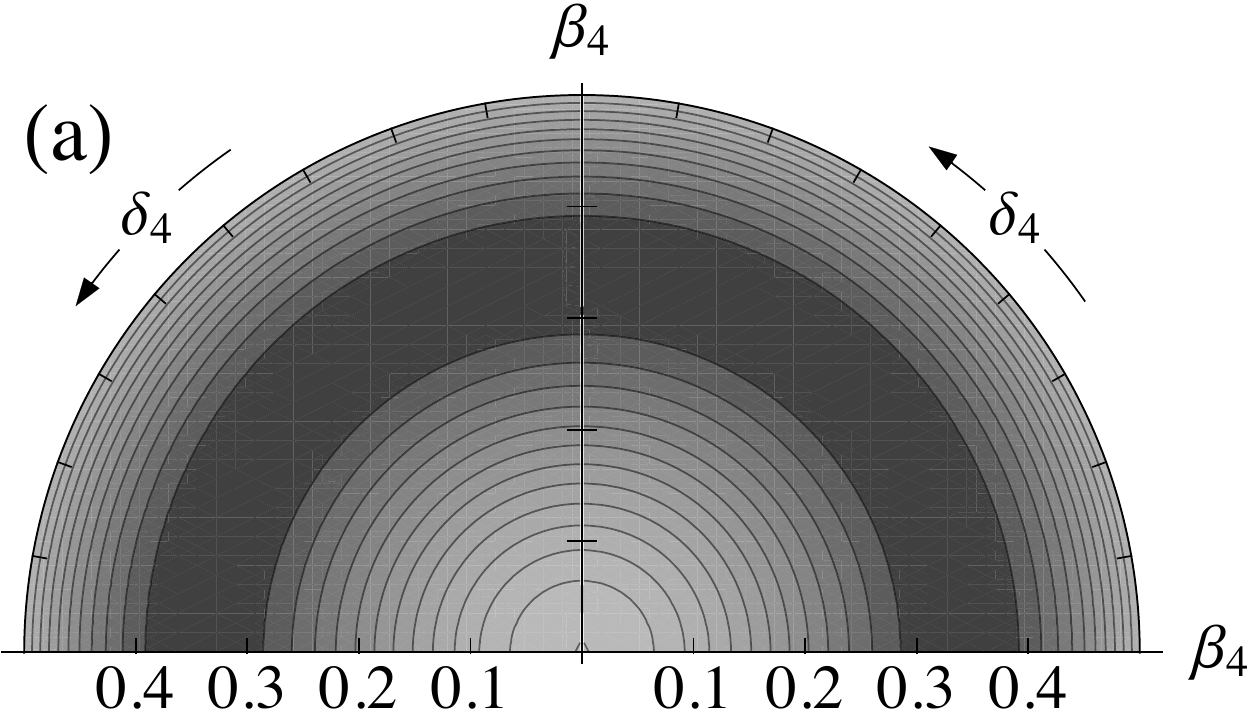}
\includegraphics[width=6cm]{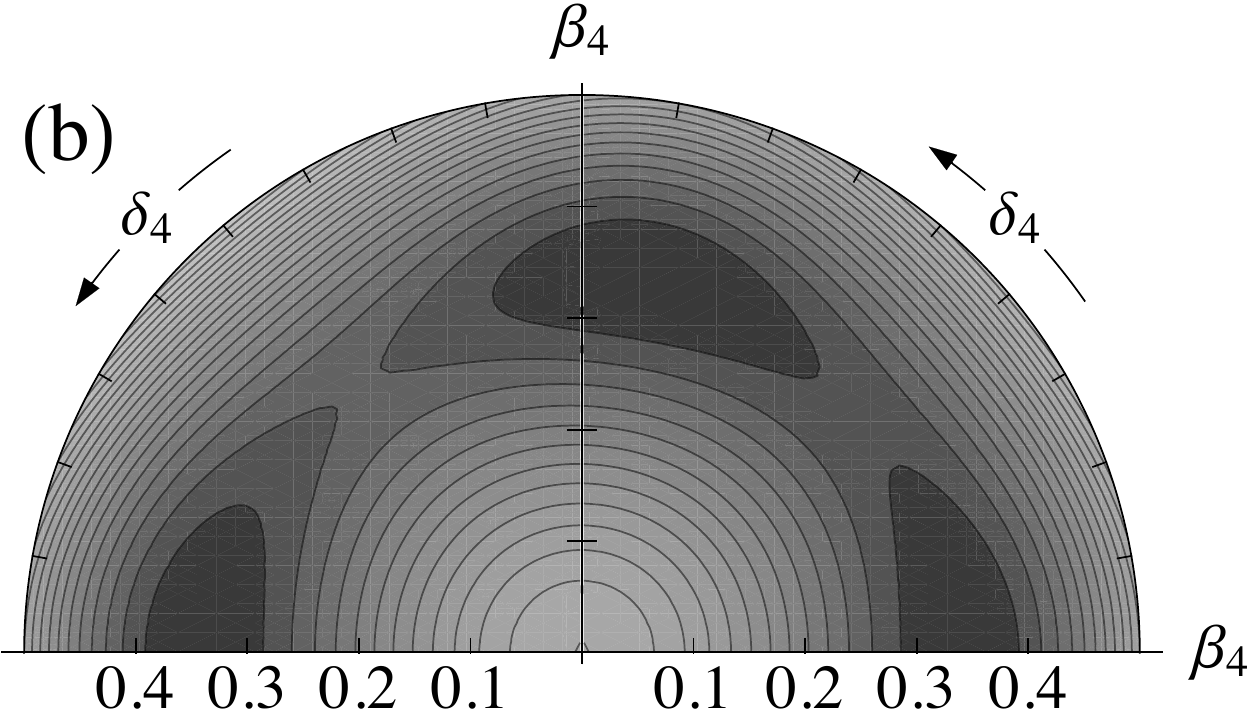}
\caption{Energy surfaces $E(\beta_4,\delta_4)$
obtained in the classical limit
of two different Hamiltonians of the \mbox{$sdg$-IBM} for $N=5$ bosons.
The dependence on $\beta_4>0$ and $0\leq\delta_4\leq\pi$ is shown for $\beta_2^*=0$.
Black corresponds to low energies
and the lines indicate changes by 10~keV.
(a) The ${\rm U}_g(9)$--${\rm SO}_{sg}(10)$ transitional Hamiltonian is taken
with the parameters given in the caption of Fig.~\ref{f_u9so10}.
(b) The Hamiltonian of (a) is modified by taking a repulsive interaction $v_{gggg}^2=500$~keV.}
\label{f_b4d4}
\end{figure}
The parameters quoted in the caption of Fig.~\ref{f_u9so10}
satisfy both conditions~(\ref{e_cond}).
As a result, the energy surface in the classical limit of the corresponding Hamiltonian
has a minimum for $\beta_2^*=0$ and $\beta_4^*\approx0.34$,
as shown in Fig.~\ref{f_b2b4}.
According to the preceding discussion,
the surface is independent of $\delta_4$,
which is indeed confirmed by Fig.~\ref{f_b4d4}(a).
If the $v_{gggg}^2$ matrix element is modified,
a dependence in $\delta_4$ is introduced,
as illustrated in Fig.~\ref{f_b4d4}(b) for the value $v_{gggg}^2=500$~keV.
It is seen that the energy surface displays {\em three} isolated minima
that are exactly degenerate.
The three minima all have an octahedral symmetry,
corresponding to either an octahedron [$\delta_4^*=0$ and $\delta_4^*=\arccos(1/6)\approx84.4^{\rm o}$]
or a cube ($\delta_4^*=\pi$).

\begin{figure}
\centering
\includegraphics[width=3cm]{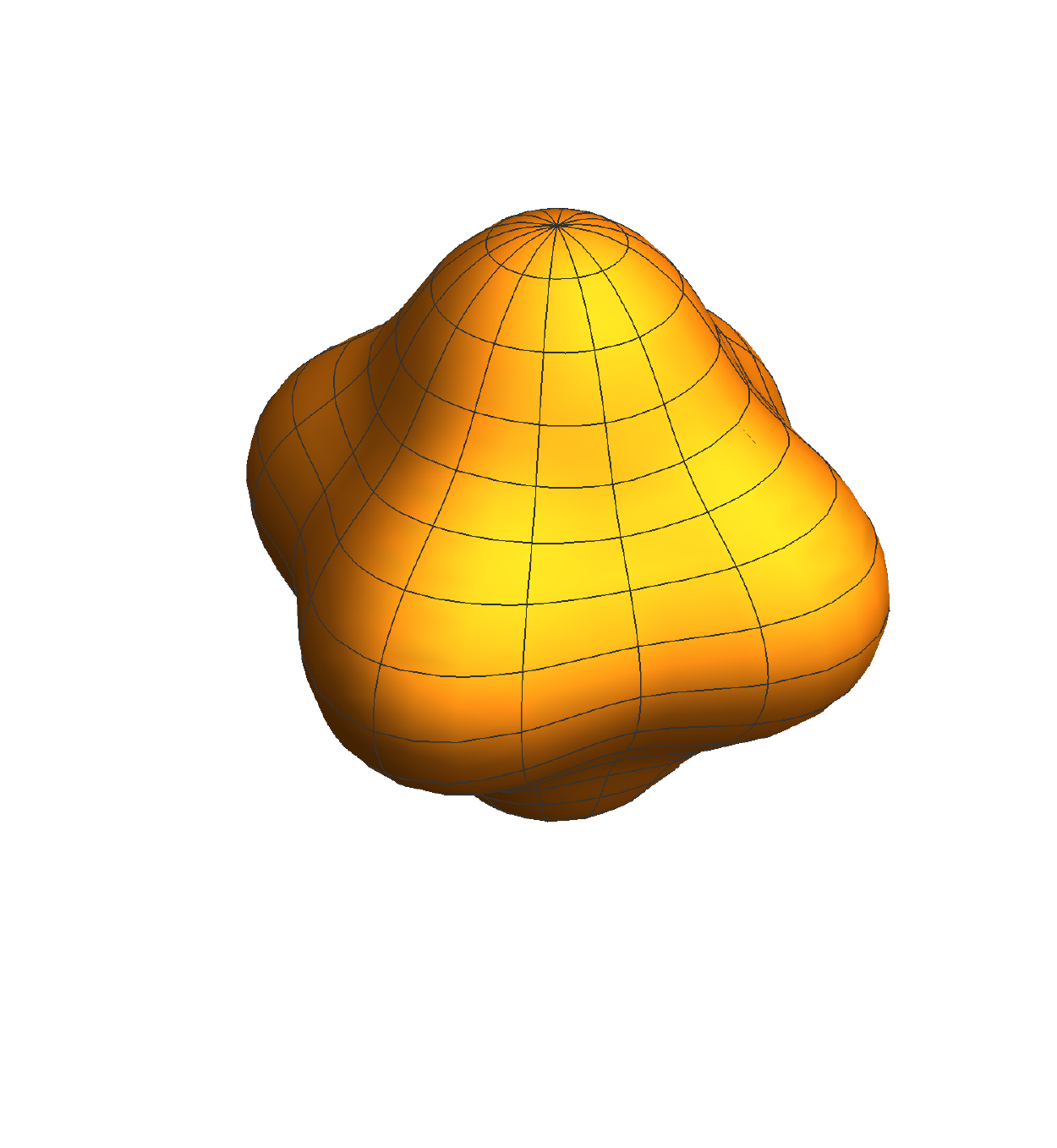}
\includegraphics[width=7cm]{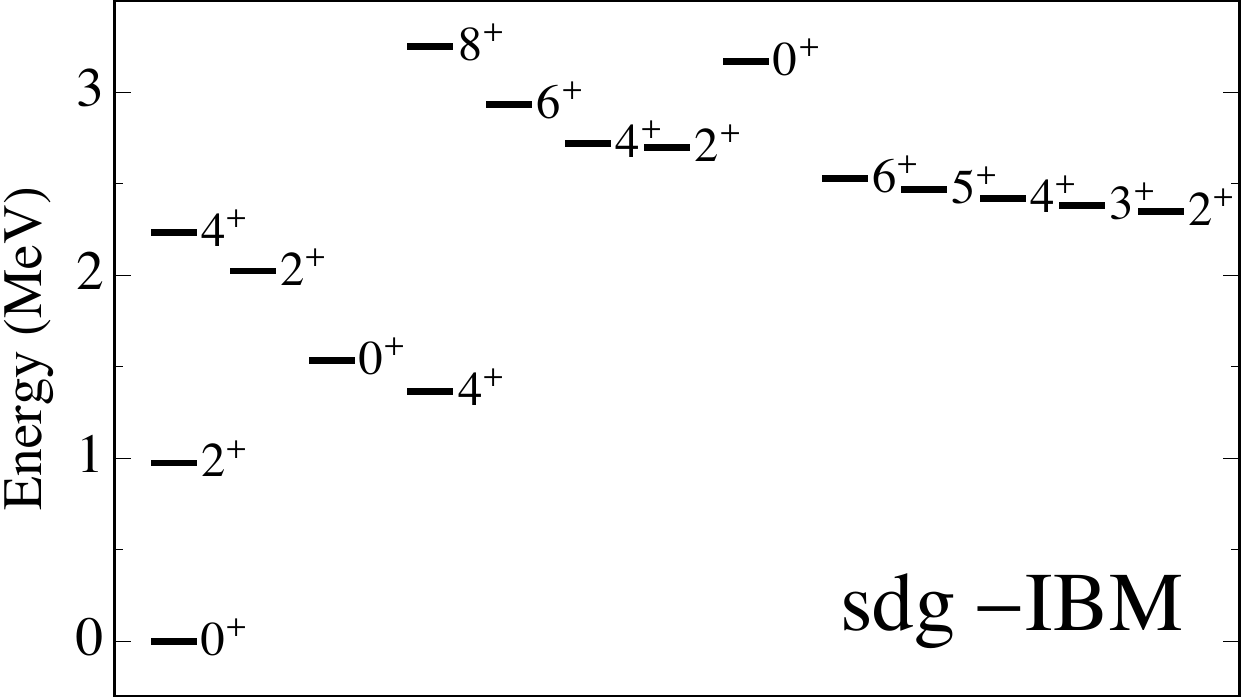}
\includegraphics[width=3cm]{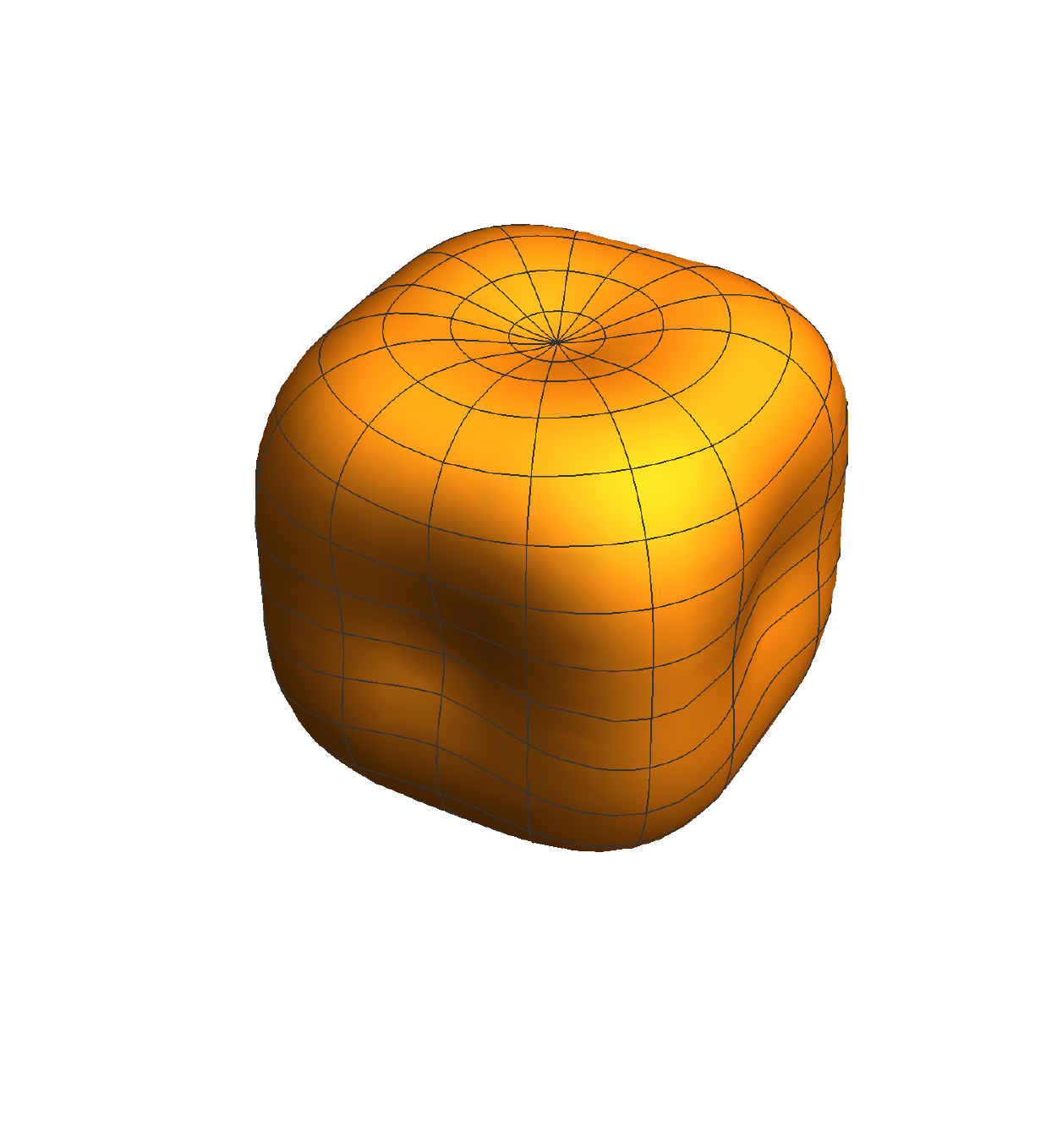}
\caption{Energy spectrum of a general Hamiltonian of the \mbox{$sdg$-IBM} for $N=5$ bosons.
The same Hamiltonian is taken as in Fig.~\ref{f_u9so10}
but one $g$-boson two-body matrix element is modified to $v_{gggg}^2=500$~keV.
On the left- and right-hand sides are shown
the shapes at the minima in the energy surface
obtained in the classical limit of this Hamiltonian.
They have octahedral symmetry and correspond to either an octahedron or a cube.}
\label{f_u15}
\end{figure}
Although this analysis shows that
isolated minima with octahedral symmetry can be obtained 
in the classical limit of an \mbox{$sdg$-IBM} Hamiltonian with reasonable parameters,
it can be expected that such minima are rather shallow.
Even for the fairly large value of the interaction matrix element
in the above example, $v_{gggg}^2=500$~keV,
the three minima are separated by a barrier of $\sim20$~keV,
inducing only very weak observable effects.
This point is illustrated with Fig.~\ref{f_u15},
which shows the spectrum of the ${\rm U}_g(9)$--${\rm SO}_{sg}(10)$ transitional Hamiltonian
with the modified $v_{gggg}^2$ matrix element.
Except for some minute changes the spectrum is essentially the same
as that shown in Fig.~\ref{f_u9so10}.

One subtle point made clear by the current study is that
it is not sufficient to carry out a catastrophe analysis
of the generic surface~(\ref{e_climit})
obtained in the classical limit of the most general \mbox{$sdg$-IBM} Hamiltonian~(\ref{e_ham})
with up to two-body interactions between the bosons.
The coefficients $c'_{kl}$ and $c^{ij}_{kl}$ cannot be treated as free parameters
but their expressions in terms of the single-bosons energies
and boson--boson interactions are an essential part of the analysis.
To illustrate this point consider the energy surface shown in Fig.~\ref{f_b4d4}(b).
The minima at $\delta_4^*=0$ and $\delta_4^*\approx84.4^{\rm o}$
correspond to the same intrinsic shape
(an octahedron, shown on the left-hand side of Fig.~\ref{f_u15})
and, as a consequence, the minima {\em must} be exactly degenerate.
This behavior is generally valid.
Therefore, whatever single-boson energies and boson--boson interactions
one adopts in the Hamiltonian~(\ref{e_ham}),
the energy surface in its classical limit must satisfy the constraint
that points on the surface with same intrinsic shape
(e.g., $\delta_4^*=0$ and $\delta_4^*\approx84.4^{\rm o}$)
are at the same energy.

\section{Conclusions}
\label{s_conc}
The main conclusion of this paper is
that no \mbox{$sdg$-IBM} Hamiltonian with a dynamical symmetry
that includes Casimir operators of up to second order
displays in its classical limit an isolated minimum with octahedral shape.
Nevertheless, a degenerate minimum that {\em includes} a shape with octahedral symmetry
can be obtained from a Hamiltonian transitional between two limits.
In the limits in question the $d$ boson is decoupled from $s$ and $g$ bosons.
Furthermore, limit I, ${\rm U}_g(9)$, has hexadecapole vibrational characteristics
while in limit II, ${\rm SO}_{sg}(10)$, $s$- and $g$-boson states
are mixed through an $sg$-pairing interaction.
A catastrophe analysis of the energy surface
obtained in the classical limit of this transitional symmetry Hamiltonian
indicates that a minimum with zero quadrupole and non-zero hexadecapole deformation
can be obtained with reasonable parameters.
However, this minimum is always $\delta_4$ independent,
meaning that it ranges from an octahedron to a cube
and includes intermediate shapes without octahedral symmetry. 
Isolated minima with octahedral symmetry can be obtained
by adding two-body interactions between the $g$ bosons
to the transitional symmetry Hamiltonian.
The resulting energy surface displays in this case minima with octahedral symmetry,
with the shape of either an octahedron or a cube,
separated by a barrier with low energy
even for fairly strong interactions between the $g$ bosons.
The conclusion of this analysis in the context of the \mbox{$sdg$-IBM}
is therefore that it will be difficult to find
experimental manifestations of octahedral symmetry in nuclei. 

\section*{Acknowledgements}
This work has been carried out
in the framework of a CNRS/DEF agreement,
project N 13760.

\end{document}